\documentclass{article}
\usepackage{ijcai21}

\usepackage{times}
\usepackage{soul}
\usepackage{url}
\usepackage[hidelinks]{hyperref}
\usepackage[utf8]{inputenc}
\usepackage[small]{caption}
\usepackage{subfigure}
\usepackage{graphicx}
\usepackage{diagbox}
\usepackage{amsmath}
\usepackage{amsthm}
\usepackage{booktabs}
\usepackage{algorithm}
\usepackage{algorithmic}
\title{Towards a Very Large Scale Traffic Simulator for Multi-Agent Reinforcement Learning Testbeds}

\author{
Zijian Hu$^1$
\and
Chengxiang Zhuge$^2$
\and
Wei Ma$^{*1}$
\affiliations
$^1$Department of Civil and Environmental Engineering, \\ The Hong Kong Polytechnic University, Kowloon, Hong Kong SAR\\
$^2$Department of Land Surveying and Geo-Informatics, \\ The Hong Kong Polytechnic University, Kowloon, Hong Kong SAR\\
\emails
zijian.hu@connect.polyu.hk,
chengxiang.zhuge@polyu.edu.hk,
wei.w.ma@polyu.edu.hk
}

\begin{document}

\maketitle

\begin{abstract}
  Smart traffic control and management become an emerging application for Deep Reinforcement Learning (DRL) to solve traffic congestion problems in urban networks. Different traffic control and management policies can be tested on the traffic simulation. Current DRL-based studies are mainly supported by the microscopic simulation software ({\em e.g.}, SUMO\footnote{\url{https://www.eclipse.org/sumo/}}), while it is not suitable for city-wide control due to the computational burden and gridlock effect. To the best of our knowledge, there is a lack of studies on the large-scale traffic simulator for DRL testbeds, which could further hinder the development of DRL. In view of this, we propose a meso-macro traffic simulator for very large-scale DRL scenarios. The proposed simulator integrates mesoscopic and macroscopic traffic simulation model to improve efficiency and eliminate gridlocks. The mesoscopic link model simulates flow dynamics on roads, and the macroscopic Bathtub model depicts vehicle movement in regions. Moreover, both types of models can be hybridized to accommodate various DRL tasks. This creates portals for mixed transportation applications under different contexts. The result shows that the developed simulator only takes 46 seconds to finish a 24-hour simulation in a very large city with 2.2 million vehicles, which is much faster than SUMO. Additionally, we develop a graphic interface for users to visualize the simulation results in a web explorer. In the future, the developed meso-macro traffic simulator could serve as a new environment for very large-scale DRL problems.

\end{abstract}

\section{Introduction}
Traffic congestion becomes one of the most severe urban problems in recent years, and smart and effective control and management strategies ({\em e.g.}, signal control, congestion pricing, ramp metering, route guidance, etc.) are in great need to alleviate the congestion issue. Traditionally, mathematical programming is employed to obtain the optimal policy on small-scale networks, while it may fail in city-wide networks due to the exponentially growing problem scale and complexity. To this end, Deep Reinforcement Learning (DRL) emerges as one of the tools for decision-makings in a large-scale and complex environment, and its application on smart traffic control and management attracts wide attentions in recent years.

The development of DRL for traffic management and control requires accurate and large-scale traffic simulators as training and testing environments. Table \ref{tab:review_RL_sim} presents a summary of traffic simulators for DRL tasks. Most simulators leverage car-following models and lane-change models in a microscopic level. However, due to the computational cost of microscopic model, it is difficult to operate a very large-scale scenario based on these simulators. Mesoscopic and Macroscopic traffic models were developed in the transportation community to improve the simulation efficiency. Mesoscopic simulation models the the congestion dynamics on links. However, the gridlock effect, a special case of traffic congestion that vehicles are blocked in a circle queue,  are prone to happen under large-scale scenarios due to the inproper settings of link attributes; macroscopic models prevent the vehicle gridlock, but it sacrifices the simulation precision. Therefore, a model that incorporates advantages of both mesoscopic models and macroscopic models would be the optimal one to fulfill the very large-scale traffic simulation.


\begin{table}[h]
  \centering
  \begin{tabular}{p{0.27\columnwidth}p{0.27\columnwidth}p{0.3\columnwidth}}
      \hline
      {\bf Simulator} & {\bf Purpose} & {\bf Scale} \\
      \hline
      CarRacing (Gym) \cite{OpenAI_Gyms} & Autonomous driving & $\leq 10$ links and nodes \\
      \hline
      Highway-env \cite{highway-env} & Autonomous driving & $\leq 10$ links and nodes \\
      \hline
      Flow \cite{Flow} & Autonomous driving & $\leq 10$ links and nodes \\
      \hline
      SMARTS \cite{SMARTS} & Autonomous driving & $\leq 10$ links and nodes \\
      \hline
      BARK \cite{BARK} & Behavior modeling & $\leq 10$ links and nodes \\
      \hline
      CityFlow \cite{CityFlow} & Signal control & 2,510 nodes and 25,156 vehicles \cite{Thousand_Lights} \\
      \hline
      {\bf This paper} &  {\bf Multi-purpose} & {\bf 27,000 nodes, 80,000 links, 2.2 million vehicles} \\
      \hline
  \end{tabular}
  \caption{A review of different traffic simulators for reinforcement tasks.}
  \label{tab:review_RL_sim}
\end{table}


In this paper, we propose a meso-macro level traffic simulator for city-wide control and management, and we integrate the regional and link-based vehicle dynamic models as the backbone models. To our best knowledge, this is the first traffic simulator designed for very large-scale reinforcement learning testbeds. We adopted the traffic scenario from Turin, Italy \cite{TUST} as the benchmark for evaluation, which contains about 27,000 nodes and 80,000 links with 2.2 million vehicles in a day. The developed simulator only takes about 46 seconds to finish 24-hour simulation on a single thread program. To visualize the simulation result, we provide both vehicle trajectories and link volumes using JavaScript and Python. The developed meso-macro traffic simulation could serve as a new environment for very large-scale DRL problems.

\section{Proposed Work}
\subsection{Design and Structure}
\begin{figure}[H]
  \centering
  \includegraphics[width=0.40\textwidth]{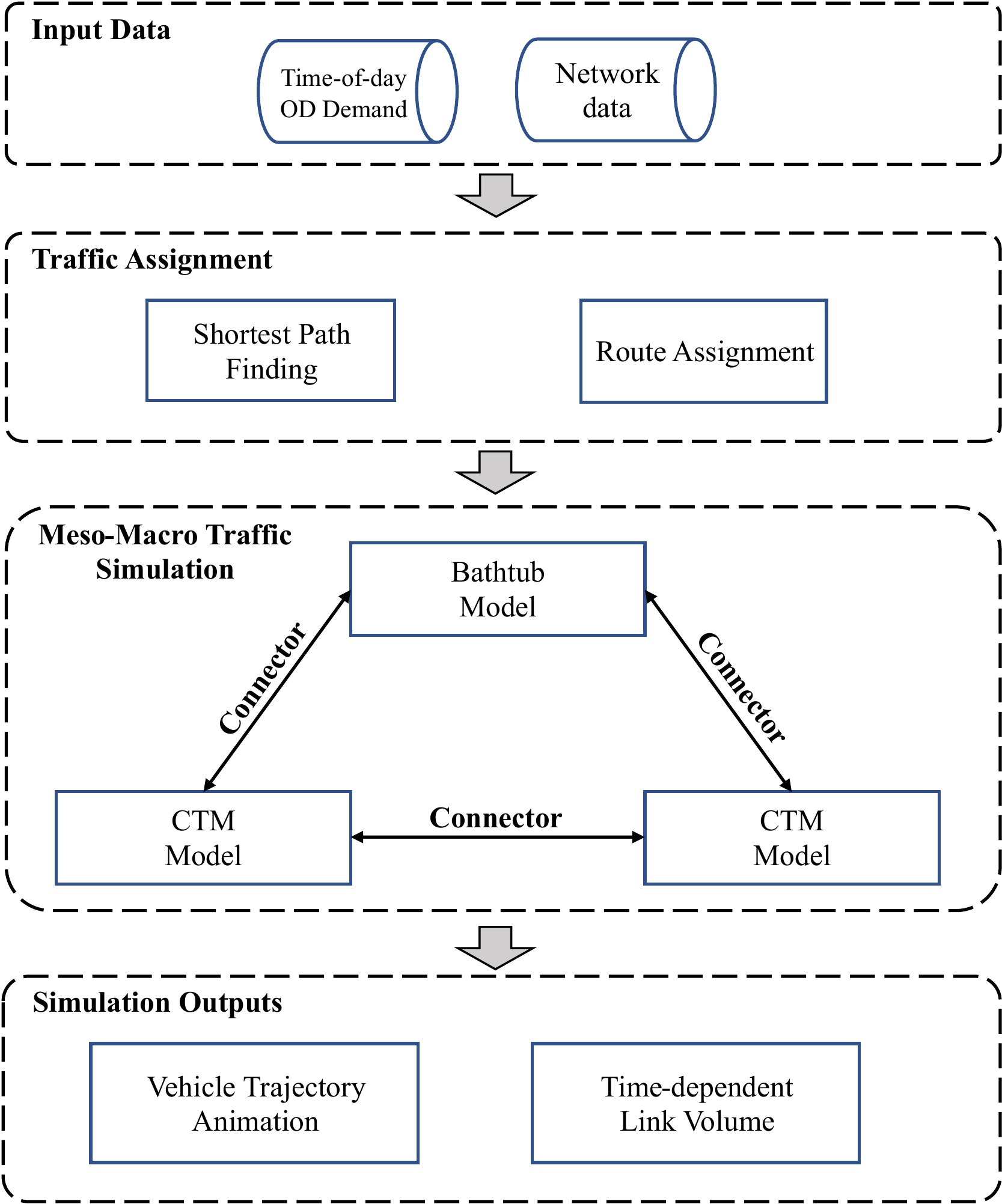}
  \caption{The structure of proposed simulator.}
  \label{fig:framework}
\end{figure}

The meso-macro traffic simulator models link dynamics and vehicle behaviors at each time step, providing fruitful information that can be used for DRL tasks. The framework of our simulator is shown in Figure~\ref{fig:framework}, which consists of three major components: traffic assignment, simulation, and simulation outputs.

\begin{itemize}
  \item The Traffic Assignment module aims to assign a path for each vehicle based on road network data and time-of-day Origin-Destination (OD) pairs data.
  \item Meso-Macro Traffic Simulation module incorporates three traffic models, \textit{Generalized Bathtub Model} \cite{Bathtub}, \textit{Cell Transmission Model} (CTM) \cite{CTM1,CTM2} and \textit{Link Transmission model} (LTM) \cite{LTM}. These models can be hybridized, and the corresponding connectors will be developed to regulate the entrance and exit of vehicles between each two models.
  \item The Simulation Output module can generate the vehicle trajectory and time-dependent link volumes on the networks.
\end{itemize}

\subsection{Traffic Assignment}
Traffic assignment module takes the information of network and time-dependent OD data to provide different paths for vehicles based on the link travel time. We prepare two initial algorithms for the assignment module, All-or-Nothing (AON) assignment and Incremental assignment. AON method assigns vehicles to the shortest path under the free-flow travel time. Drivers with the same OD will share the same route at any time. Incremental assignment fractures the total vehicles into pieces. In each step, a piece of vehicles is assigned base on the current shortest paths. The process will be repeated until all vehicles are assigned to specific paths.

\subsection{Meso-Macro Traffic Simulation}
The meso-macro traffic simulation module is the core component of the developed simulator, where vehicles are driven by different traffic models. The operation of vehicles can be mainly separated into two types, in-link operation and between-link operation. For in-link operations, vehicles evolve according to different traffic models. For example, the travel time of vehicle on CTM or LTM can be inferred from the link length and density at each time, while in the Bathtub model, the travel time of vehicles are determined by the remaining distance and vehicle numbers in the same region. For vehicle operation between links, we developed a node model and six connectors for vehicle transfer between different models. The detailed information of models and connectors lists as follows.

\subsubsection{Cell Transmission Model}
The CTM discretizes the link into segments, and vehicles are transfered between different segments, which simulates the movement of vehicles on roads. The length of each segment is defined as the travel distance at free-flow speed in a time interval, formulated as $\Delta x = v_f \Delta t$, where $v_f$ denotes the free-flow speed. The conservation law of inflow and outflow vehicles in each segment can be stated as
\begin{equation}
  n_i^j(t) = n_i^j(t - \Delta t) + \gamma_i^j(t) - \gamma_{i+1}^j(t),
  \nonumber
\end{equation}
\noindent where $n_i^j(t)$ represents the vehicle number in segment $i$ on link $j$ at time $t$. $\gamma_i^j(t)$ denotes the number of inflow vehicles from segment $i-1$ on link $j$ at time $t$, and $\gamma_{i+1}^j(t)$ means the number of outflow vehicles from segment $i$ to segment $i+1$ on link $j$ at time $t$.

The inflow or outflow number is determined by the sending flow or receiving flow as
\begin{equation}
  \begin{aligned}
    \gamma_i^j &= \min \{ S_{i-1}^j, R_i^j \} \\
    S_{i-1}^j &= \min \{ k_{i-1}^j(t - \Delta t) v_f^j \Delta t, q_m^j \Delta t \} \\
    R_i &= \min \left \{ q_m^j \Delta t , \left( K^j - k_i^j(t - \Delta t)v_b^j \Delta t \right) \right \},
  \end{aligned}
 \nonumber
\end{equation}
\noindent where $S_{i-1}^j$ represents the maximum sending vehicles from segment $i-1$, and $R_i$ denotes the maximum receiving flow in segment $i$. $k_{j-1}(t - \Delta t)$ represents the traffic density in previous segment (the unit is vehicle per lane per kilometer), $q_m^j$ represents the maximum traffic flow that it can undertake (the unit is vehicle per second). $K^j$ denotes the jam density and $w_b^j$ denotes the spillback speed of congestion.

\subsubsection{Link Transmission Model}
The LTM records the movement of vehicles by maintaining a physical queue for each link, rather than transferring vehicles in segments in CTM. 
Supposing $n_0^j(t)$ and $n_L^j(t)$  represent vehicle numbers passing throw the origin and destination of link $j$ at time $t$. The maximum sending and receiving vehicles are defined as
\begin{equation}
  \begin{aligned}
    S^j &= \min \left \{ q_0^j \left(t_f^j \right) \Delta t + n_0^j \left( t_f^j \right) - n_L^j(t), q_m^j \Delta t \right \} \\
    R^j &= \min \left \{ q_L^j \left(t_b^j \right) \Delta t + n_L^j \left( t_b^j \right) + K^jL^j - n_0^j(t), q_m^j \Delta t \right \} \\
    t_f^j &= t - \frac{L^j}{v_f^j} \\
    t_b^j &= t - \frac{L^j}{v_b^j},
  \end{aligned}
 \nonumber
\end{equation}
\noindent where $q_0^j(t)$ and $q_L^j(t)$ denote the traffic flow at upstream and downstream of link $j$ at time $t$.

\subsubsection{Bathtub Model}
The Bathtub model assumes a relationship between vehicle speed and regional vehicle numbers, which is not constrained by the jam density and maximum flow in CTM and LTM. We use {\bf boldface} font to differentiate symbols from CTM and LTM models. 
The vehicle speed $\mathbf{v}(t)$ is defined as
\begin{equation}
    \mathbf{v}(t) = V\left( \frac{\lambda(t)}{\mathbf{L}} \right) \quad \text{and} \quad 
    \lambda(t) = \sum_{i=0}^{\mathbf{L}/\Delta x} \mathbf{k}(t, i\Delta x),
\nonumber
\end{equation}
\noindent where $\lambda(t)$ denotes the total vehicle number in a region. $\mathbf{k}(t, x)$ denotes the vehicle numbers with uniform remaining distance of $x$ at time $t$. $\mathbf{L}$ represents the longest path length in this region. The space-time dependent vehicle number $\mathbf{k}(t, x)$ can be updated as:
\begin{equation}
  \begin{aligned}
      \mathbf{k}_{r, s}^{p, i}(t + \Delta t, x) & =
      \mathbf{k}_{r, s}^{p, i}\left( t, x + \mathbf{v}_i(t) \Delta t \right)  
      \\ &+ \sum_{\forall j \in \Gamma^-(i)} z \left(x, \mathbf{L}_{r, s}^{p, i} \right) \cdot \mathbf{k}_{r, s}^{p, j}(t, 0) \\ & + 
      z \left( x, \mathbf{L}_{r, s}^{p, r} \right) \cdot \mathbf{k}_{r, s}^{p, r} \left( t, \mathbf{L}_{r, s}^{p, r} \right),
  \end{aligned}
  \nonumber
\end{equation}

\noindent where  $\mathbf{k}_{r, s}^{p, i}(t, x)$ denotes the number of vehicles that begins from region $r$, ends to region $s$, chooses path $p$ in region $i$. $\Gamma^-(i)$ represents the upstream regions to region $i$, and $z(\cdot, \cdot)$ is the XNOR function. Note that in the CTM and LTM, all vehicles obey the First-in-First-out (FIFO) principle, meaning that rear vehicles can not exceed front vehicles in the same link. In the Bathtub model, FIFO is only enforced for each path separately. 

\subsubsection{Node Models and Connectors}
\begin{figure}[h] \centering    
  \subfigure[An example of node model in CTM and LTM.] {
   \label{fig:link_node_model}     
  \includegraphics[width=0.4\columnwidth]{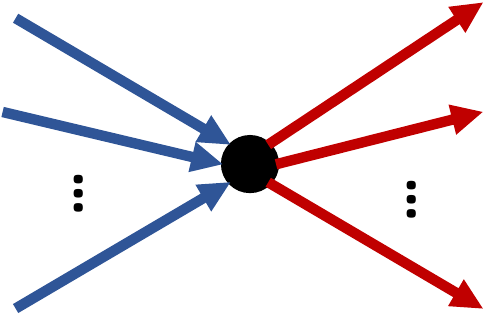}  
  }
  \quad
  \subfigure[An example of node model in Bathtub model.] {
   \label{fig:region_node_model}     
  \includegraphics[width=0.3\columnwidth]{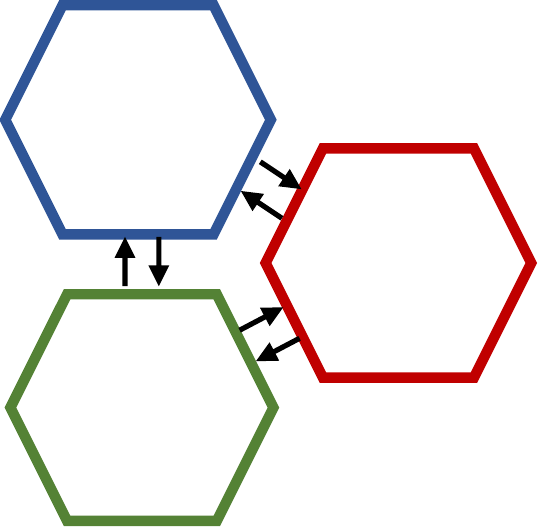}  
  }
  \subfigure[An example of connectors.] { 
  \label{fig:node_model}     
  \includegraphics[width=0.8\columnwidth]{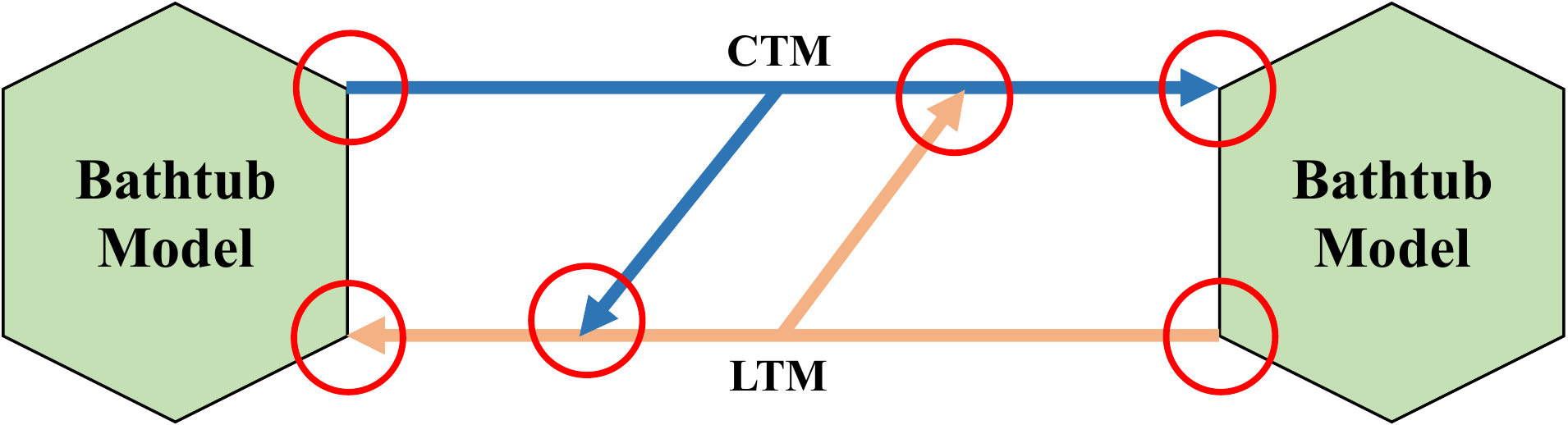}     
  }
  \caption{The illustration of node models and connectors}
  \label{fig:node_model_connectors}
  \end{figure}

Node models and connectors regulate the vehicle transfer between links with same and different models, as shown in Figure~\ref{fig:node_model_connectors}. The basic idea behind these two modules are the same: for each node (in CTM or LTM) and region (in Bathtub Model), we randomly select a previous link or region and check the following link or region of the top vehicle. If the following link or region can accommodate a vehicle, we move the vehicle to the new link or region. Otherwise, the vehicle should wait until there is a vacancy. A noteworthy detail is that when the unit of the vehicle is just a little bit larger than the vacancy, the vehicle will be split into two pieces and one will be sent to the following link or region. The rest will wait until the next time step. For example, the vehicle unit is 1 and the remaining vacancy is 0.6, the vehicle will be split into 0.4 and 0.6. We let the one with 0.6 enter and keep the other till the next time step.

\subsection{Simulation Outputs}
For the simulation outputs, we provide a web-based graphic interactive animation. Users can check the speed, volume and trajectory of each vehicle after simulation. Since the large-scale rendering is an exhausting computation for the web explorer. We utilize a high-efficient WebGL-powered framework of geospatial data visualization \textit{AntV L7}\footnote{\url{https://l7.antv.vision}} for large-scale simulation rendering. The link volumes, travel time, and vehicle trajectories can be further summarized to generate rewards for different DRL tasks.

\begin{figure*}[t]
  \centering    
  \subfigure[Turin map from OSM] {
   \label{fig:google_map}     
  \includegraphics[width=0.66\columnwidth]{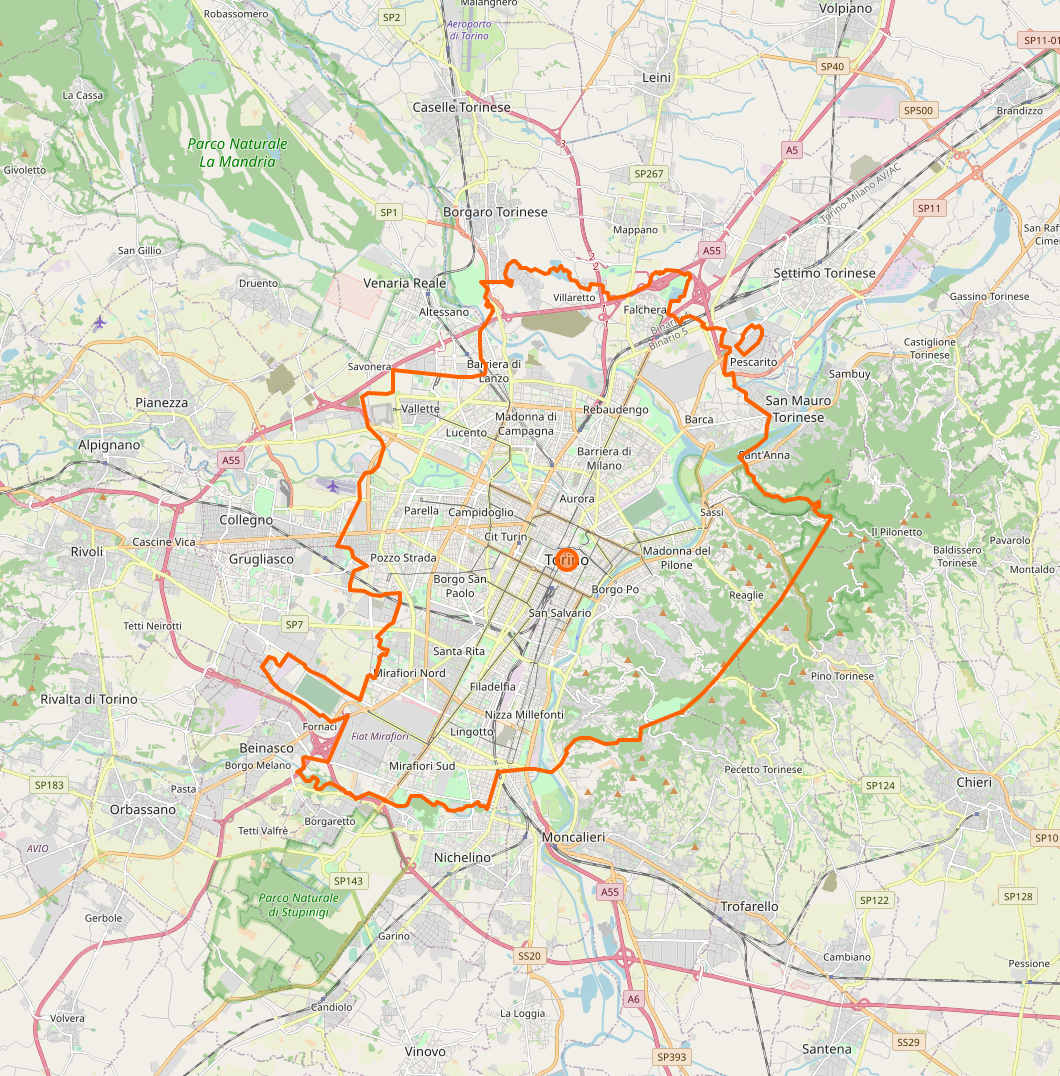}  
  }
  \subfigure[Simplified network topology] { 
  \label{fig:network_edges}     
  \includegraphics[width=0.6\columnwidth]{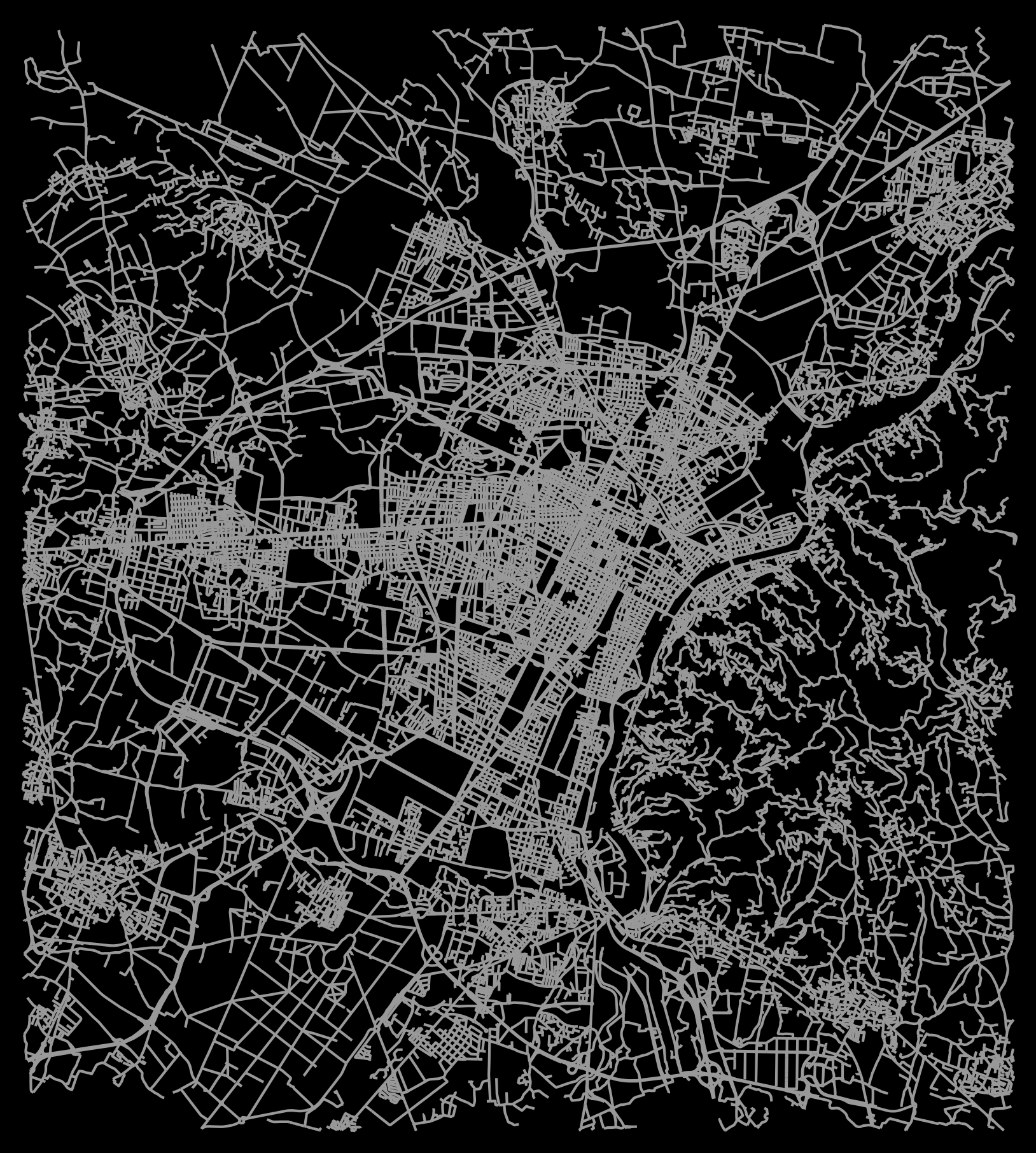}     
  }
  \subfigure[Network clustering] { 
  \label{fig:network_clusters}     
  \includegraphics[width=0.6\columnwidth]{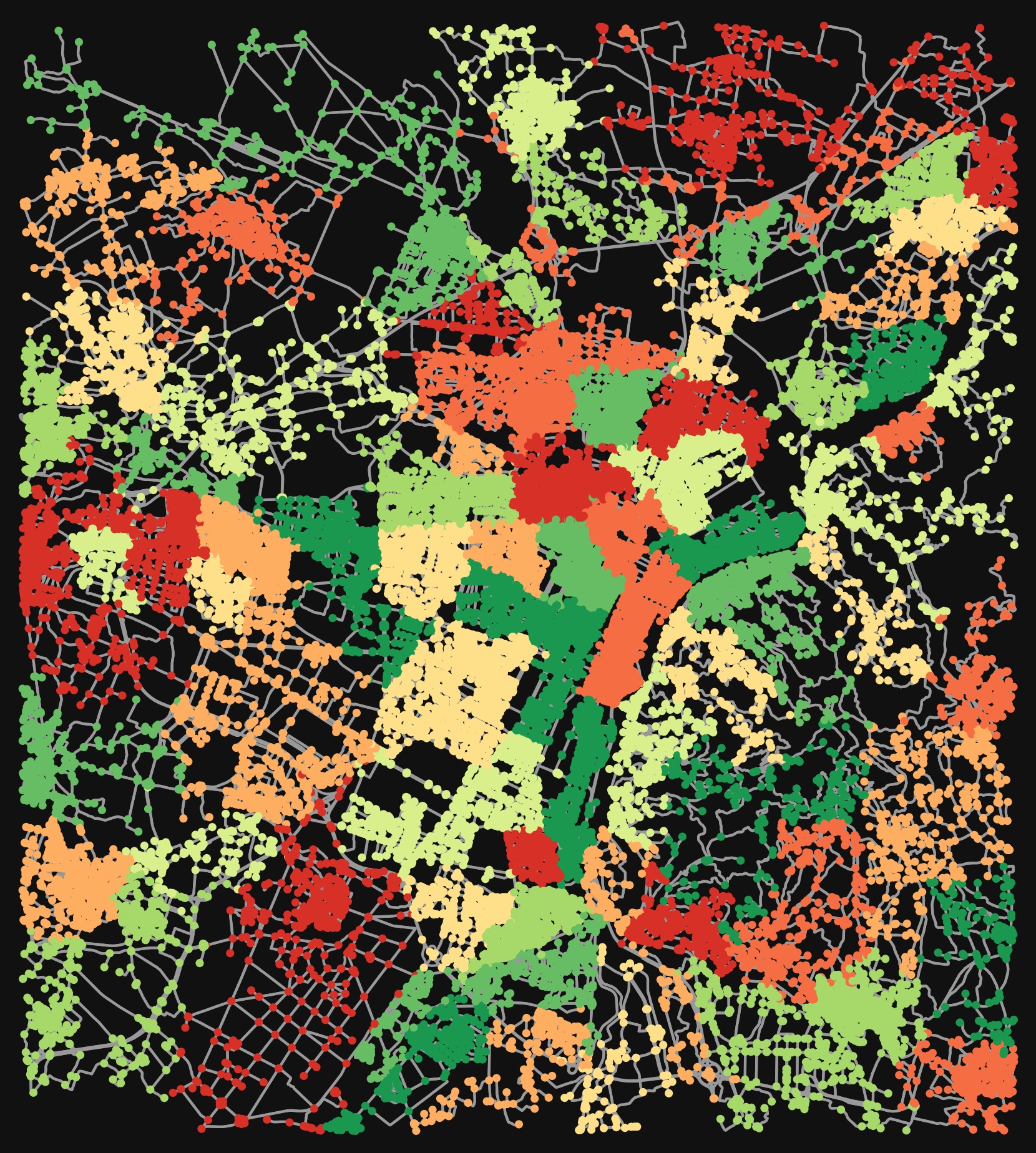}     
  }
  \caption{Turin maps and community clustering.}
  \label{fig:networks}     
\end{figure*}

\section{Experiments}

\subsection{Scenarios}
To evaluate the simulation performance in a large-scale network, we take advantage of a SUMO scenario called TuST in Turin, Italy \cite{TUST}. The road network covers about 600 square kilometers with 32,936 nodes and 66,296 links. This scenario contains 24-hour OD demand data with a total number of 2.2 million vehicles stemmed from real traffic data in Turin. According to our best knowledge, this is the largest public scenario that we can have so far.

\subsection{Implementation details}
\subsubsection{Network Creation}
The network was automatically created by using a Python package \textit{OSMnx} \cite{OSMnx}, which can convert the map from the OpenStreetMap into MultiDiGraph class in Networkx (shown in Figure~\ref{fig:google_map} and \ref{fig:network_edges}). OSMnx can also correct the broken links and remove non-junctional nodes, which improves the efficiency of the simulation. And the final node and edge numbers are 27,231 and 79,063. Since not all edges contain speed limit information, links without speed limit are imputed by a hard-code table in SUMO according to road type and lane number. The whole simulation is implemented by C++ for efficiency consideration.

\subsubsection{Network Partition}
To partition the road network into homogeneous regions, we utilize the Leiden algorithm, which is a community detection algorithm \cite{Leiden}, to clusters nodes into different regions. For Turin network, 84 different regions are finally divided and each region at least contains 100 nodes (shown in Figure~\ref{fig:network_clusters}). The Underwood's model is calibrated to depict the relationship between speed and number of vehicles for each region \cite{MFD_Calibration}. 


\subsubsection{Experimental Settings}
All experiments are divided into two parts. In the first part, we compare the simulation performance with CTM, LTM and Bathtub model. In the second part, the performance of SUMO is compared in the same scenarios. All experiments are run on AMD Ryzen 9 5900X@4.8GHz with a memory of 64GB@2666MHz. The time interval is set to 1 second. The total demand for 24-hour is about 2.2 million vehicles and each vehicle is assigned based on the AON method.

\begin{figure}[H]
  \centering
  \includegraphics[width=0.5\textwidth]{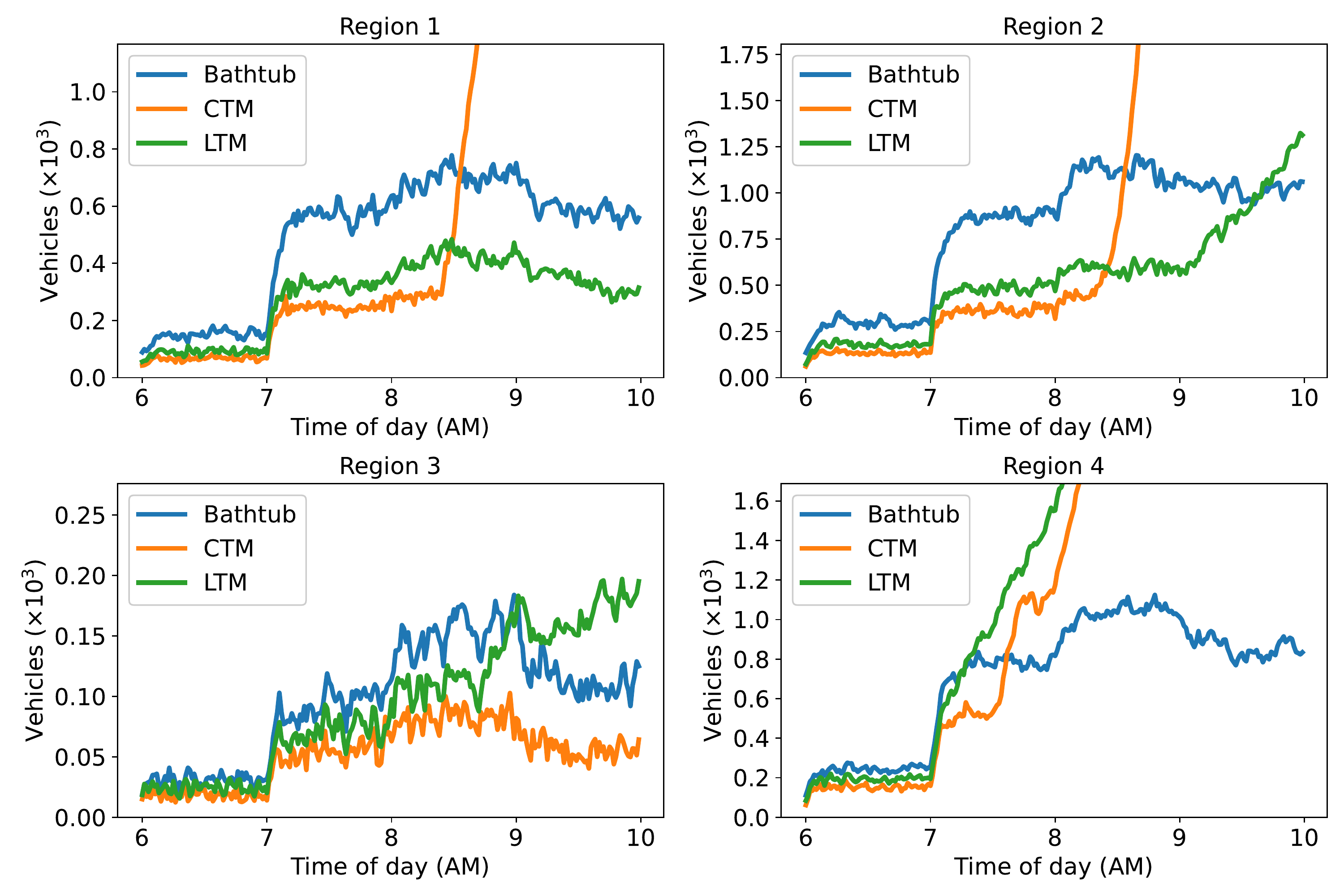}
  \caption{Regional vehicle accumulation based on different traffic models.}
  \label{fig:region_acc_0010}
\end{figure}

\subsection{Results}
\subsubsection{Comparison with CTM, LTM and Bathtub}
\begin{figure}[h]
  \centering
  \includegraphics[width=0.48\textwidth]{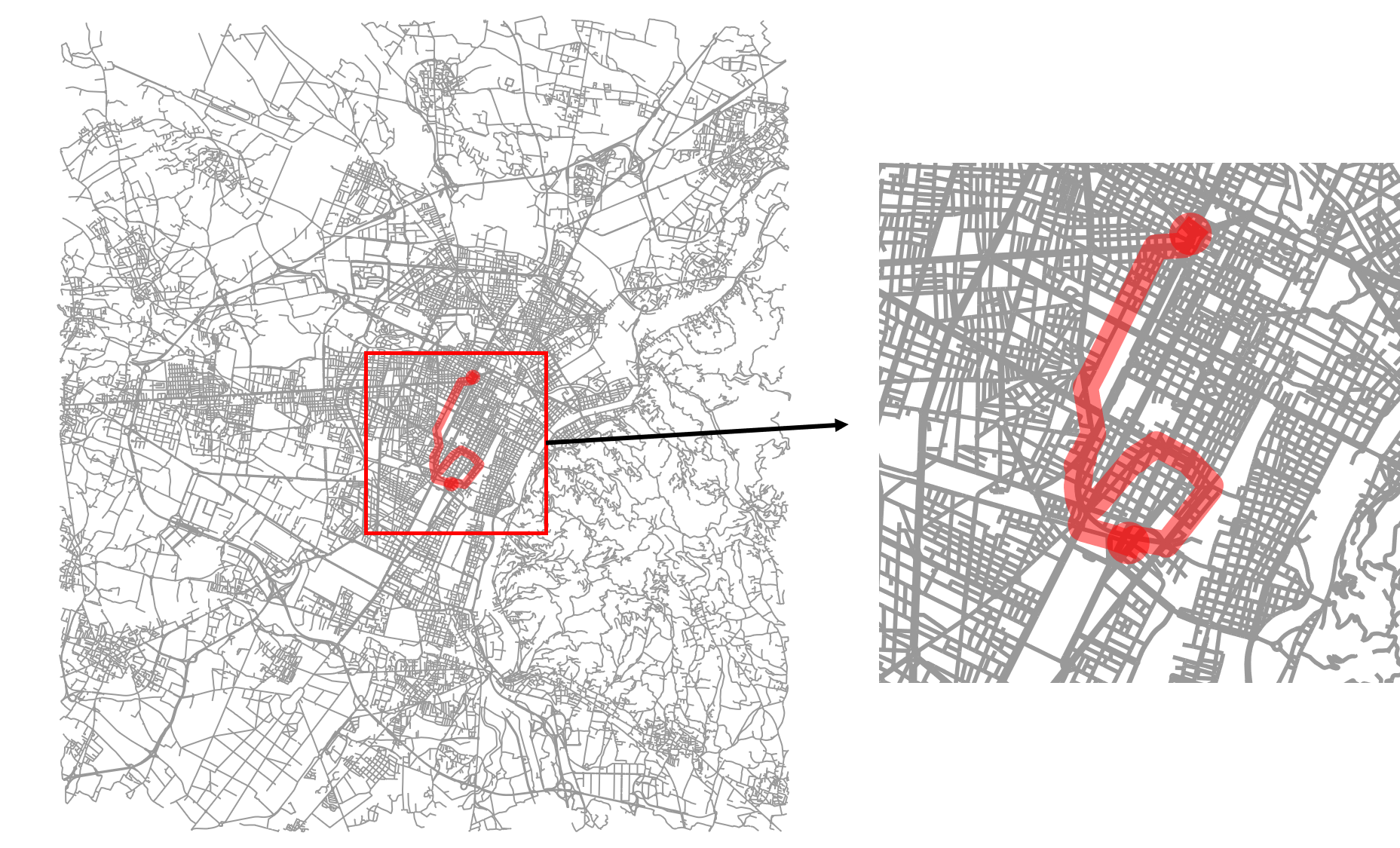}
  \caption{A case of gridlock in simulation.}
  \label{fig:gridlock}
\end{figure}

In this section, we compared the results based on different traffic models. Since the gridlock effect frequently appear in CTM and LTM models, making it impossible to finish the simulation, even just for the morning peak, we only compare the results from 6 AM to 10 AM. Having partitioned the network into 84 homogeneous regions, Figure~\ref{fig:region_acc_0010} shows the vehicle accumulation in the first four regions. It is clear that there is a dramatic increase for vehicles in Region 1, 2 and 4 based on the CTM model and Region 2, 3, 4 in LTM models also demonstrate gridlock effects. Figure~\ref{fig:gridlock} shows an entity of gridlock based on CTM model, in which all the vehicles cannot move for more than an hour. The gridlock effect begins with circle congestion since all entrances and exits are blocked by vehicles. Eventually, gridlock will propagate along the upstream road and paralyze the whole network finally. Though the vehicle accumulation is slightly higher based on the Bathtub model, it overcame this problem, making it possible to simulate a very large-scale network since the Bathtub model does not rely on link capacities or speed limit.

\subsubsection{Comparison with Bathtub, Hybrid and SUMO}
In this section, we mainly compared the performance between the proposed simulator and SUMO. We set up four groups of comparison with three scenarios in each group: 1) fully Bathtub model; 2) hybrid model, 82 regions for the Bathtub model and 2 regions for CTM models; 3) fully CTM; 4) SUMO. The demand scale is fixed to three levels: 20\%, 100\% and 200\%. In the hybrid model, regions for CTM models are manually selected where the average length of links in a region is long. The duration of experiments has been elongated to 24 hours to evaluate the consumption of each model. Table~\ref{tab:simulation_time} shows the simulation time within different scenarios. For the 20\% demand, the simulation time in all groups is in a reasonable time interval. Given the normal demand situation, it nearly costs four hours for CTM models and SUMO due to the gridlock effect even if we enable the auto-correction of gridlock in SUMO, while the Bathtub model only takes 46 seconds to finish the whole simulation, much more efficient than the CTM model and SUMO. And if we double the demand to 4.4 million vehicles per day, it only takes about 2 minutes to finish the whole simulation, while the CTM and SUMO are not tested considering the gridlock effect on normal demand scenarios.
\begin{table}[h]
  \centering
  \begin{tabular}{|l|p{0.13\columnwidth}|p{0.10\columnwidth}|p{0.10\columnwidth}|p{0.10\columnwidth}|}
      \hline
      \diagbox{Demands}{Time (s)}{Model} & Fully Bathtub & Hybrid & Fully CTM & SUMO \\
      \hline
      20\% & 19 & 521 & 1,359 & 384 \\
      \hline
      100\% & 46 & 669 & 11,987 & 13,567 \\
      \hline
      200\% & 135 & 976 & \diagbox[innerwidth=0.10\columnwidth, dir=SW]{} & \diagbox[innerwidth=0.10\columnwidth, dir=SW]{} \\ \\
      \hline
  \end{tabular}
  \caption{Simulation time under different scenarios within 24 hours.}
  \label{tab:simulation_time}
\end{table}

\begin{figure}[h]
  \centering
  \includegraphics[width=0.5\textwidth]{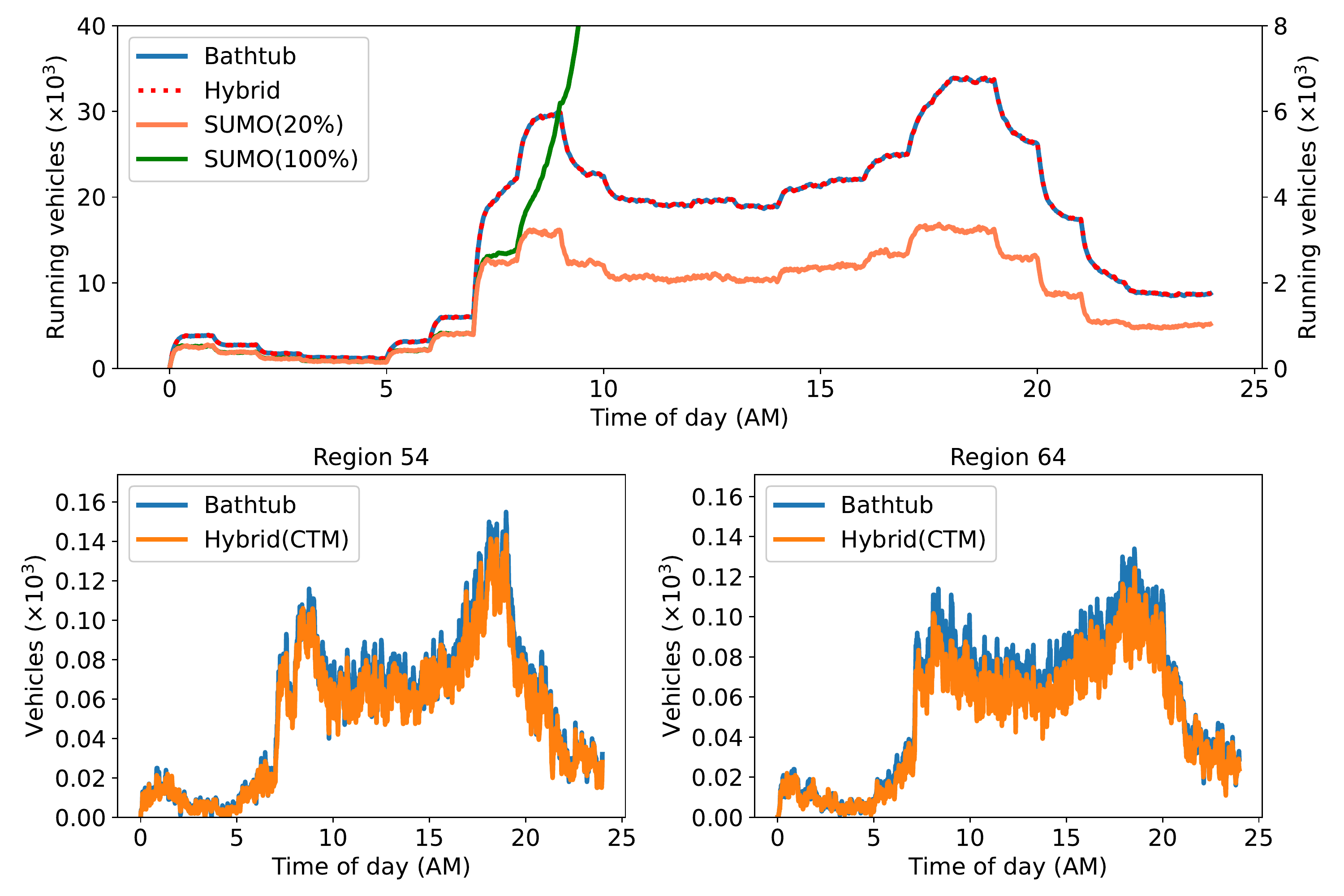}
  \caption{Vehicle accumulation based on different scenarios.}
  \label{fig:partial_acc_allday}
\end{figure}

Figure~\ref{fig:partial_acc_allday} presents the vehicle accumulation with different scenarios in the whole network within 24 hours. Note that the scale of 20\% demand SUMO group corresponds with the right vertical axis while the other three groups correspond with the left vertical axis. The lower two pictures indicate the major influence on regional vehicle number may come from the in-flow and out-flow rate of the region. The excessive in-flow rate will corrupt the stability of the whole system by gridlock eventually. The result also demonstrates that the Bathtub model can overcome the gridlock effect during large-scale traffic simulation. The regional vehicle number based on the Bathtub and hybrid model can simulate more than four times of the vehicles than SUMO.


\subsubsection{Visualization Demo}
\begin{figure}[h]
  \centering    
  \subfigure[Vehicle trajectory animation] {
    \label{fig:vehicle_traj}
    \includegraphics[width=0.47\textwidth]{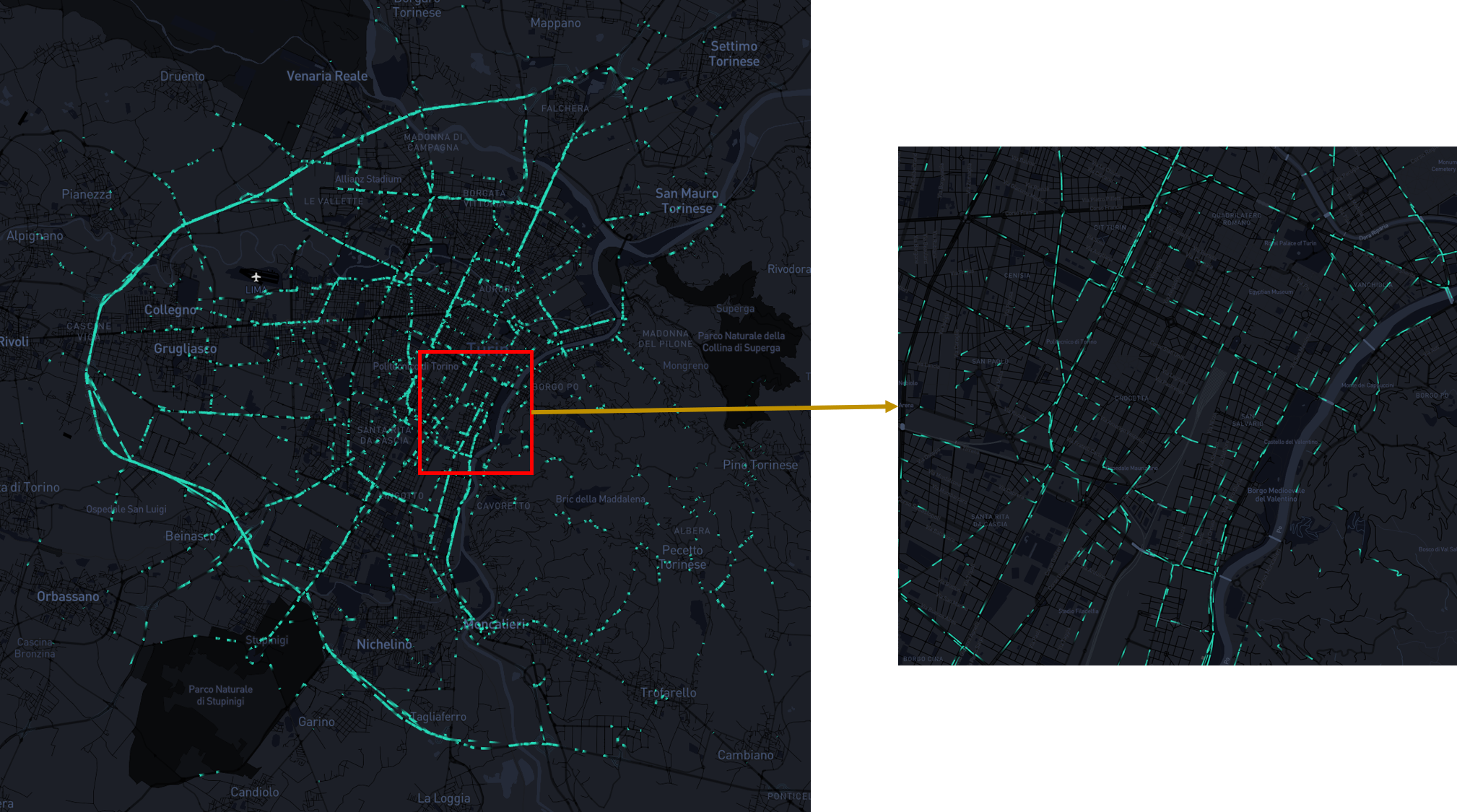}
  }
  \subfigure[Link volume variation] {
    \label{fig:link_volume}
    \includegraphics[width=0.47\textwidth]{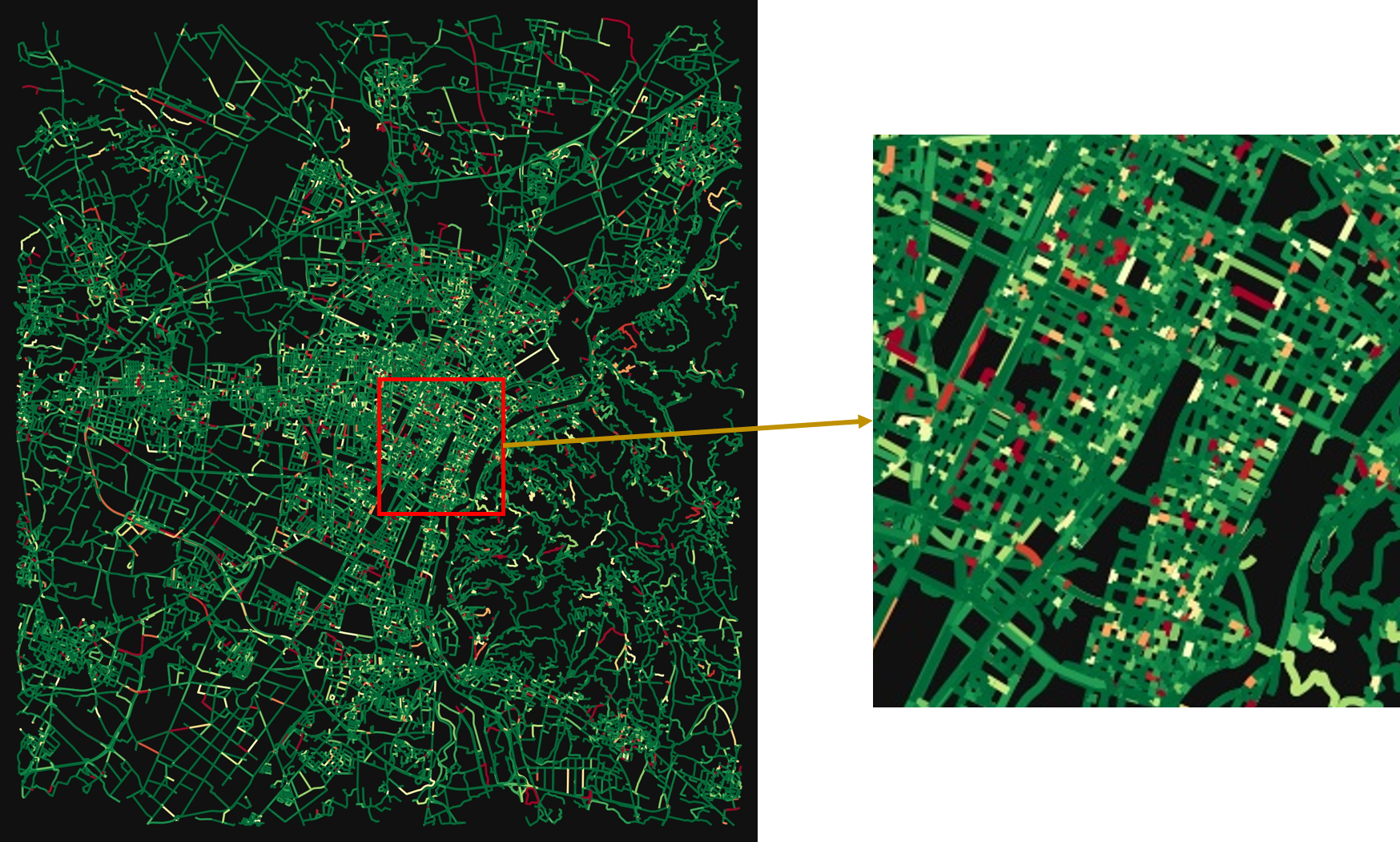}
  }
  \caption{A demo of visualization.}
  \label{fig:visualization_demo}     
\end{figure}

A demo of a visualization is shown in Figure~\ref{fig:visualization_demo}. Figure~\ref{fig:vehicle_traj} shows a trajectory animation with 3,000 vehicles. The green line segments represent simulated vehicles in the urban network. The longer the segment is, the faster the vehicle drives. Figure~\ref{fig:link_volume} shows the traffic density in each link where red means that link is congested and green means link is empty. We are still actively developing the visualization platform to support more options for checking simulation states.

\section{Discussion and Future works}
This section presents three potential DRL tasks of the Intelligent Transportation System (ITS) using the proposed simulator. 
With the increase of network scale, the dimension and complexity of the DRL task grows exponentially, making it nearly impossible to obtain analytical solutions in a constant time. DRL method has been demonstrated to solve high-dimension and large-scale traditional control tasks in a constant time. Moreover, the Bathtub model can achieve a highly efficient simulation, making it possible for training DRL applications. It is a good attempt to manage urban mobility using the DRL method under the context of very large-scale networks. 
To summarize, the following three potential tasks for traffic controls using DRL methods are proposed.

\subsection{Corridor Management}
Corridor management aims to controls the mobility in major transportation corridors based on the information from ITS \cite{ICM}. It is crucial to ease the congestion problem by controlling the inflow and outflow in Central Business District (CBD). Typically, it is suitable for cities with heterogeneous traffic demand. Taking metropolitan areas such as San Francisco as an instance, a newborn family is likely to choose suburban areas considering the housing price, while most schools and office buildings are concentrated in the downtown area. The commuting time on workdays might be prolonged, and hence the traffic condition is especially vital for commuters in mega-cities.
The synchronous management for multiple corridors may happen based on the proposed network. The populated areas ({\em e.g.}, households, office buildings, school, etc.)  can be modeled as homogeneous regions, simulated under the Bathtub model, while CTM or LTM can model corridors. Observations, actions and rewards can be delicately designed to match the requirement for DRL.

\subsection{Vehicle Rerouting Problem}
Vehicle Rerouting Problem mainly focuses on minimizing the utility cost facing unexpected traffic accidents. When accidents happen, how will the others react to minimize the influence? The problem can be divided into two parts, in-accident-region rerouting and out-accident-region rerouting. In-accident-region rerouting aims to guide vehicles to leave the impacted area as soon as possible. Due to the character of hybrid simulation of our simulator, we can flexibly change the impact regions into the link-based model and output the optimal rerouting policy for each vehicle using the DRL. When the accident has been solved by police, we change the region back to the Bathtub model again to maintain a high-efficient simulation. For out-region rerouting problems, they may need to change regional paths to react to sudden inflow caused by accident. Two sub-problems can be treated as a cooperation game and can be solved by DRL methods.

\subsection{Jointly management with Signal Control and Perimeter Control}
Signal control is one of the earliest tasks that leverage DRL to solve the congestion problem \cite{PressLight}. Currently, thousands of traffic signals can be controlled using the DRL under microscopic traffic models \cite{Thousand_Lights}. It works when the total travel demand is small. When it comes to a 24-hour simulation, it nearly impossible to finish one step due to the computational burden and gridlock effect. The Bathtub simulation has been proved to be insensitive to the gridlock effect and high efficiency. However, it cannot be directly applied for signal control since it cannot model the spillback of congestion. An alternative solution is to achieve perimeter control in different regions based on the Bathtub model, but it can only reach coarse-granular management for the urban network. Therefore, a compatible solution is to achieve an intermediate-resolution control by joint management of signal and perimeter. For those regions we are interested in, we can achieve a fine-granular control by manipulating signals. And for those we are not, we can apply for perimeter controls. The unified scenario can be regarded as a DRL task under the ITS context.

\section{Conclusion}
In this paper, we proposed a meso-macro traffic simulator for very large-scale network scenarios. The simulator integrates three mesoscopic and macroscopic traffic models: CTM, LTM, and the Bathtub model. Users can achieve a flexible combination of models in simulation with different purposes. We evaluate the performance of the developed simulator and compare with the prevailing simulator SUMO. The experiments have demonstrated that the gridlock effect in the simulation can be avoided, and the simulation speed is faster than SUMO in a very large-scale network, making it suitable for a DRL testbed for various traffic control and management tasks. In the future, gym-fashion environment will be implemented and open-sourced, and we will develop a more functional graphic interface for users and prepare more tasks for the implementation of DRL algorithms.

\section*{Acknowledgments}
The work described in this study was supported by the Smart Cities Research Institute (CDA9), Research Institute for Sustainable Urban Development (BBWF) and a start-up grant (BE3Q) at the Hong Kong Polytechnic University. 

\clearpage

\bibliographystyle{named}
\bibliography{ijcai21}

\begin{thebibliography}{}

\bibitem[\protect\citeauthoryear{Bernhard \bgroup \em et al.\egroup
  }{2020}]{BARK}
Julian Bernhard, Klemens Esterle, Patrick Hart, and Tobias Kessler.
\newblock {BARK}: Open behavior benchmarking in multi-agent environments.
\newblock In {\em 2020 IEEE/RSJ International Conference on Intelligent Robots
  and Systems (IROS)}, 2020.

\bibitem[\protect\citeauthoryear{Boeing}{2017}]{OSMnx}
Geoff Boeing.
\newblock {OSMnx}: New methods for acquiring, constructing, analyzing, and
  visualizing complex street networks.
\newblock {\em Computers, Environment and Urban Systems}, 65:126--139, 2017.

\bibitem[\protect\citeauthoryear{Brockman \bgroup \em et al.\egroup
  }{2016}]{OpenAI_Gyms}
Greg Brockman, Vicki Cheung, Ludwig Pettersson, Jonas Schneider, John Schulman,
  Jie Tang, and Wojciech Zaremba.
\newblock Openai gym.
\newblock {\em arXiv preprint arXiv:1606.01540}, 2016.

\bibitem[\protect\citeauthoryear{Chen \bgroup \em et al.\egroup
  }{2020}]{Thousand_Lights}
Chacha Chen, Hua Wei, Nan Xu, Guanjie Zheng, Ming Yang, Yuanhao Xiong, Kai Xu,
  and Zhenhui Li.
\newblock Toward a thousand lights: Decentralized deep reinforcement learning
  for large-scale traffic signal control.
\newblock {\em Proceedings of the AAAI Conference on Artificial Intelligence},
  34(04):3414--3421, Apr. 2020.

\bibitem[\protect\citeauthoryear{Daganzo}{1994}]{CTM1}
Carlos~F. Daganzo.
\newblock The cell transmission model: A dynamic representation of highway
  traffic consistent with the hydrodynamic theory.
\newblock {\em Transportation Research Part B: Methodological}, 28(4):269--287,
  1994.

\bibitem[\protect\citeauthoryear{Daganzo}{1995}]{CTM2}
Carlos~F. Daganzo.
\newblock The cell transmission model, part ii: Network traffic.
\newblock {\em Transportation Research Part B: Methodological}, 29(2):79--93,
  1995.

\bibitem[\protect\citeauthoryear{Jin}{2020}]{Bathtub}
Wen-Long Jin.
\newblock Generalized bathtub model of network trip flows.
\newblock {\em Transportation Research Part B: Methodological}, 136:138--157,
  2020.

\bibitem[\protect\citeauthoryear{Klim \bgroup \em et al.\egroup }{2016}]{ICM}
Terry Klim, Anna Giragosian, Diane Newton, Elisabeth Bedsole, and Robert
  Sheehan.
\newblock Integrated corridor management, transit, and mobility on demand.
\newblock Technical report, United States. Federal Highway Administration,
  2016.

\bibitem[\protect\citeauthoryear{Leurent}{2018}]{highway-env}
Edouard Leurent.
\newblock An environment for autonomous driving decision-making.
\newblock \url{https://github.com/eleurent/highway-env}, 2018.

\bibitem[\protect\citeauthoryear{Rapelli \bgroup \em et al.\egroup
  }{2019}]{TUST}
Marco Rapelli, Claudio Casetti, and Giandomenico Gagliardi.
\newblock {TuST}: From raw data to vehicular traffic simulation in turin.
\newblock In {\em 2019 IEEE/ACM 23rd International Symposium on Distributed
  Simulation and Real Time Applications (DS-RT)}, pages 1--8, Oct 2019.

\bibitem[\protect\citeauthoryear{Traag \bgroup \em et al.\egroup
  }{2019}]{Leiden}
V.~A. Traag, Ludo Waltman, and Nees Jan~van Eck.
\newblock From {Louvain} to {Leiden}: guaranteeing well-connected communities.
\newblock {\em Scientific Reports}, 9(1):5233, 2019.

\bibitem[\protect\citeauthoryear{Wei \bgroup \em et al.\egroup
  }{2019}]{PressLight}
Hua Wei, Chacha Chen, Guanjie Zheng, Kan Wu, Vikash Gayah, Kai Xu, and Zhenhui
  Li.
\newblock Presslight: Learning max pressure control to coordinate traffic
  signals in arterial network.
\newblock In {\em Proceedings of the 25th ACM SIGKDD International Conference
  on Knowledge Discovery; Data Mining}, KDD '19, page 1290–1298, New York,
  NY, USA, 2019. Association for Computing Machinery.

\bibitem[\protect\citeauthoryear{Wong \bgroup \em et al.\egroup
  }{2021}]{MFD_Calibration}
Wai Wong, S.~C. Wong, and Henry~X. Liu.
\newblock Network topological effects on the macroscopic fundamental diagram.
\newblock {\em Transportmetrica B: Transport Dynamics}, 9(1):376--398, 2021.

\bibitem[\protect\citeauthoryear{Wu \bgroup \em et al.\egroup }{2017}]{Flow}
Cathy Wu, Aboudy Kreidieh, Kanaad Parvate, Eugene Vinitsky, and Alexandre~M
  Bayen.
\newblock Flow: A modular learning framework for autonomy in traffic.
\newblock {\em arXiv preprint arXiv:1710.05465}, 2017.

\bibitem[\protect\citeauthoryear{Yperman}{2007}]{LTM}
Isaak Yperman.
\newblock {\em The Link Transmission Model for Dynamic Network Loading}.
\newblock PhD thesis, Katholieke Universiteit Leuven, 01 2007.

\bibitem[\protect\citeauthoryear{Zhang \bgroup \em et al.\egroup
  }{2019}]{CityFlow}
Huichu Zhang, Siyuan Feng, Chang Liu, Yaoyao Ding, Yichen Zhu, Zihan Zhou,
  Weinan Zhang, Yong Yu, Haiming Jin, and Zhenhui Li.
\newblock {CityFlow}: A multi-agent reinforcement learning environment for
  large scale city traffic scenario.
\newblock In {\em The World Wide Web Conference}, WWW '19, page 3620–3624,
  New York, NY, USA, 2019. Association for Computing Machinery.

\bibitem[\protect\citeauthoryear{Zhou \bgroup \em et al.\egroup
  }{2020}]{SMARTS}
Ming Zhou, Jun Luo, Julian Villella, Yaodong Yang, David Rusu, Jiayu Miao,
  Weinan Zhang, Montgomery Alban, Iman Fadakar, Zheng Chen, Aurora~Chongxi
  Huang, Ying Wen, Kimia Hassanzadeh, Daniel Graves, Dong Chen, Zhengbang Zhu,
  Nhat Nguyen, Mohamed Elsayed, Kun Shao, Sanjeevan Ahilan, Baokuan Zhang,
  Jiannan Wu, Zhengang Fu, Kasra Rezaee, Peyman Yadmellat, Mohsen Rohani,
  Nicolas~Perez Nieves, Yihan Ni, Seyedershad Banijamali, Alexander~Cowen
  Rivers, Zheng Tian, Daniel Palenicek, Haitham bou Ammar, Hongbo Zhang, Wulong
  Liu, Jianye Hao, and Jun Wang.
\newblock {SMARTS}: Scalable multi-agent reinforcement learning training school
  for autonomous driving.
\newblock In {\em Conference on Robot Learning}, 2020.

\end{thebibliography}

\end{document}